\documentclass{llncs}
\usepackage{makeidx}  
\usepackage{textcomp}
\usepackage{multirow}
\usepackage{moreverb}
\usepackage{latexsym}
\usepackage{graphicx,subfigure}
\usepackage{times}
\usepackage{algorithmic}
\usepackage{algorithm}
\usepackage{mathtools}
\usepackage[dvipsnames]{xcolor}

\usepackage{pgfplots}

\begin{document}
\frontmatter          

\title{Enhanced Session Initiation Protocols for Emergency Healthcare Applications}
\titlerunning{Session Initiation Protocols}  
%
\author{Saha Sourav\inst{1} \and Vanga Odelu\inst{2} \and Rajendra Prasath\inst{1}}
\authorrunning{Sourav et al.} 
\institute{Indian Institute of Information Technology Sri City, Andhra Pradesh - 517646, India\\
\email{sourav.s@iiits.in and  rajendra.prasath@iiits.in},\\ 
\texttt{http://www.iiits.ac.in/old-file/dr-rajendra-prasath}
\and
Birla Institute of Technology \& Science Pilani, Hyderabad Campus, Hyderabad - 500078, India\\
\email{odelu.vanga@hyderabad.bits-pilani.ac.in},\\
\texttt{http://universe.bits-pilani.ac.in/hyderabad/odeluvanga/Profile}
}

\maketitle              

\begin{abstract}
In medical emergencies, an instant and secure messaging is an important service to provide quality healthcare services. A session initiation protocol (SIP) is an IP-based multimedia and telephony communication protocol used to provide instant messaging services. Thus, design of secure and efficient SIP for quality medical services is an emerging problem. In this paper, we first explore the security limitations of the existing SIPs proposed by Sureshkumar et al. and Zhang et al. in the literature. Our analysis shows that most of the existing schemes fail to protect the user credentials when unexpectedly the session-specific ephemeral secrets revealed to an adversary by the session exposure attacks. We then present a possible improvement over Sureshkumar et al.'s scheme without increasing the computational cost. We compare the proposed improvement for computational overheads and security features with the various related existing schemes in the literature. 

\end{abstract}
%

\section{Introduction}
With the recent advances in the mobile healthcare applications, demand for secure SIP for emergency messaging alert is dramatically increasing. The e-health services present one of the major societal and economic challenges around the world, particularly for the aging society. Due to rapid growth in the number of aged people who are suffering from chronic diseases, it is emerging to improve the fast and quality low cost healthcare services. As a result, a primary focus is shifted towards delivering real-time health monitoring and quality healthcare services to the patients from their respective localities in a secure and efficient way, particularly in the medical emergency \cite{hussain2015health}. In emergency medical services (EMS), system can send an emergency request when a patient is in a critical situation. There are several EMS available where the emergency request (instant message)/ multimedia services (transmission of voice and video calls) can be sent via the cellular networks (4G/LTE,3G) \cite{alesanco2010clinical}, \cite{thelen2016using}.

\par
In the last couple of years, Voice-over-IP (VoIP) has been used mostly for multimedia data communication. The VoIP facilitates to make calls over the standard Internet broadband connection instead of public switched telephone network \cite{islam2017provably}, \cite{goode2002voice}, \cite{mishra2016secure}. On the other hand, SIP is typically used for IP-based telephony authentication, which is robust and superior over VoIP for instant messaging, internet telephone calls as well as Internet multimedia messages. SIP is used for multimedia data communications in 4G/LTE or 3G mobile networks by the 3GPP (3G Partnership Project) \cite{farash2016security}. Primarily, SIP has been standardized by the Internet Engineering Task Force standard for IP telephony \cite{campbell2002session}. SIP is a client/server based authentication scheme and it works based on digest access authentication protocol for HTTP (Hyper Text Transport Protocol) \cite{franks1999http}. In the healthcare system, when a patient wants to send an emergency request, he/she has to perform the authentication process with remote server for secure communication. According to Salsano et al. \cite{salsano2002sip} and Keromytis et al. \cite{keromytis2012comprehensive}, it is quite facile for a malicious/unauthorized user to raise a spam call or send a manipulated message to the server. If an adversary can eavesdrop, intercept or modify the emergency request, it can be catastrophic for a patient.

\begin{figure*}[!ht]
\centering
\includegraphics[width=4.8in]{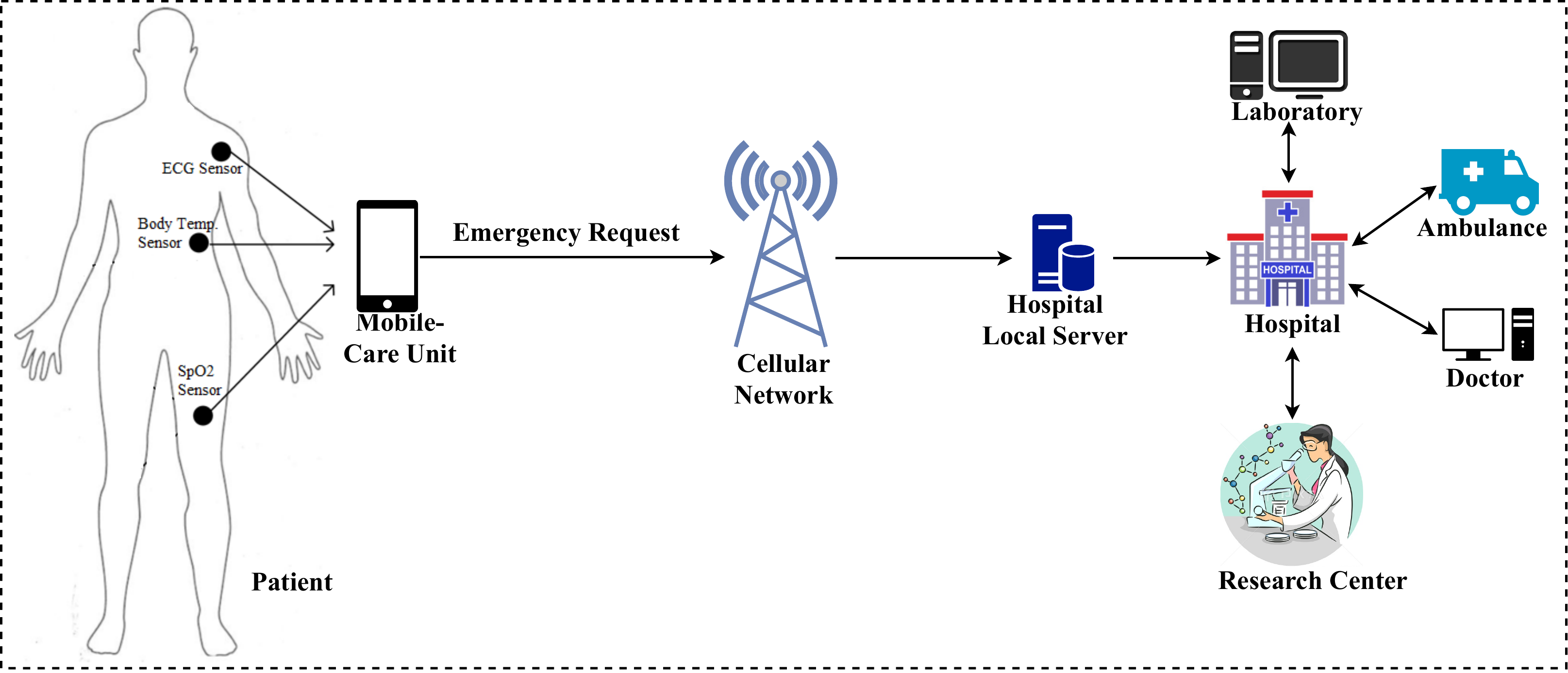}
\caption{Proposed architecture for IoT based patient monitoring System.} \label{fig-1}
\end{figure*}

We consider a network scenario  as shown in Figure \ref{fig-1}. In this model, mobile care unit (MCU) collect the information from the body sensors of patient. MCU is responsible to monitor the patient's health data and send the emergency request to the hospital server through a cellular network.  Since the health data and resources are valuable in the emergency medical situations, the hospital server (authentication server) must ensure the validity of the received request.  Therefore, ensure security while maintaining efficiency is one of the important concern in SIP for healthcare applications. In this paper, we consider the widely accepted Canetti-Krawczyk adversary (CK-adversary) model \cite{canetti2001analysis} to analyze the existing SIP. According to CK-adversary model, an authentication protocol should satisfy the following two security properties \cite{odelu2016provably}, \cite{odelu2015secure}.

\begin{itemize}
\item Future sessions should secure even if a session-specific ephemeral secrets are unexpectedly revealed to an adversary through session exposure attack.
\item All past sessions should secure even if the long-term keys of some/all the users as well as server compromise to an adversary. 
\end{itemize}
In addition, the user secret credentials should be protected against an adversary, that is, even session-specific ephemeral revealed to an adversary, he/she cannot derive the user secret credentials such identity and password.

\subsection{Organization of the paper}
The rest of the paper is organized as follows. In Section 2, we briefly discuss the required mathematical preliminaries to review and analyze the security pitfall of the existing schemes. In Section 3, we discuss the related work. In Sections 4 \& 5, we review and analyze the security weakness of the existing schemes. We then discuss the possible improvement in Section 6.  The performance analysis is presented in Section 7. Finally, we discuss the conclusion and future work in Section 8.

\section{Mathematical preliminaries}
A non-singular elliptic curve $y^2 = x^3 + ax + b$ over the finite field $GF(q)$ is the set $E_q$ of solutions $(x, y)$ $\in Z_q \times Z_q$ to the congruence $y^2 = x^3 + ax + b \, (\bmod \, q)$, where $a, b \in Z_q$ are constants such that $4a^3 + 27b^2 \neq 0 \, (\bmod \, q)$, along with the point at infinity or zero point, denoted by $\mathcal{O}$, $Z_q = \{ 0, 1, \ldots, q-1\}$ and $q > 3$ be a prime. The set of elliptic curve points $E_q$ forms an abelian group under addition modulo $q$ operation \cite{bk:01}.

\par
Let $P$ be a base point on $E_q(a,b)$ and generates a cyclic group $G$, whose order is $n$, that is, $nP = P + P + \ldots + P ( \, times) $ $= \mathcal{O}$. 

\par
The elliptic curve point multiplication is defined as the repeated additions. For example, if $P \in E_q(a,b)$, then $5P$ is computed as $5P = P + P + P + P +P \, (\bmod \, q)$.

\begin{definition}
Computing $Q = kP$ is relatively easy for given $k \in Z_q$ and $P \in G$. But, computing scalar $k$ for given $P$ and $Q = kP$ is computationally difficult problem, known as \emph{elliptic curve discrete logarithm problem (ECDLP)}. 
\end{definition}

\begin{definition}[Computational Diffie-Hellman problem (CDHP)] 
Given the parameters $P, xP, yP \in G$, computing the value $xyP \in G$ is computationally hard without the knowledge of either $x\in Z_q^{*}$ or $y \in Z_q^{*}$, where $Z_q^* =$ $\{a | 0 < a < q, \, \gcd(a, q) = 1\}$ $= \{1, 2, 3, \ldots, q-1 \}$. 
\end{definition}

\begin{definition}[Collision-resistant one-way hash function] 
A collision-resistant one-way hash function $h : X \rightarrow Y$, where $X = \{0, 1\}^{*}$ and $Y = \{0, 1\}^n$, is considered as a deterministic algorithm which takes arbitrary length input binary $x \in \{0, 1\}^*$ and outputs a fixed length binary string $y \in \{0, 1\}^n$  of length $n$ \cite{32}, \cite{33}. 
\end{definition}


\section{Related Work}
An authentication scheme should provide various aspects of security features for the SIP-based secure messaging system. The aim of an authentication protocol facilitate to the client and server to mutually authenticate each other and share a session key to communicate securely over the public channel. In 1999, Franks et al. \cite{franks1999http} derived the original SIP authentication scheme from HTTP digest authentication. Later, Yang et al. \cite{yang2005secure} found the limitations of Franks et al. \cite{franks1999http} that it fails achieve the off-line password guessing attack, server-spoofing attack, and Yang et al. \cite{yang2005secure} then proposed a new authentication scheme for SIP. In 2015, Zhang et al. \cite{zhang2015secure} presented an SIP authentication approach using ECC, and they claimed that their scheme satisfies all the required security features. However, Lu et al. \cite{lu2016secure} and Tu et al. \cite{tu2015improved} proved that the Zhang et al.'s proposed scheme \cite{zhang2015secure} is suffering from insider attack, impersonation attack, and failed to achieve strong mutual authentication. Further, Tu et al. presented an enhancement Tu et al. \cite{tu2015improved} over the Zhang et al.'s scheme. Lu et al. \cite{lu2016secure} also proposed an ECC base efficient SIP with less computation cost. However, both Farash \cite{farash2016security} and Chaudhry et al. \cite{chaudhry2017improved} analyzed and shown that Tu et al.'s scheme \cite{tu2015improved} still have security pitfalls as it failed to provide the user anonymity as well as insecure against impersonation attack. Farash \cite{farash2016security} further proposed an improved SIP, and simultaneously, Chaudhry et al. also proposed an improved SIP \cite{chaudhry2017improved}. In the recent, Lu et al. \cite{lu2017anonymous} and Chaudhry et al. \cite{chaudhry2017improved}  independently analyzed and showed that Farash's scheme \cite{farash2016security} is insecure against replay attack and impersonate attack, and failed to provide the user anonymity. Further they presented the improved versions to overcome the drawback. Recently, in 2017, Sureshkumar et al. \cite{sureshkumar2017robust} proposed an enhanced authentication scheme for SIP by pointing out the limitations in Lu et al. \cite{lu2016secure} and shown that it does not provide user anonymity, and the capability to resist user and server impersonation attacks. In this paper, we analyze and explore the security limitations of Zhang et al. \cite{zhang2015secure} and Sureshkumar et al. \cite{sureshkumar2017robust}. Our proposed security analysis is also applicable to the most of existing schemes in the literature, particularly we consider the related existing SIPs in the literature proposed by Lu et al. \cite{lu2016secure}, Tu et al. \cite{tu2015improved}, Farash \cite{farash2016security}, Chaudhry et al. \cite{chaudhry2017improved}, and Lu et al. \cite{lu2017anonymous}. We then propose a possible improvement to withstand the drawback find in the existing schemes, and discuss the future work.

\begin{table}
\small
\begin{center}
\caption{Notations used in this paper}\label{notations}
\begin{tabular}{|l | l|} \hline
U \& S & User \& Server \\ \hline
$ID_u$ and $pw_u$ & Chosen identity and password of U \\ \hline 
$E_q(a, b)$ \& P  & Elliptic curve defined over finite field $Z_q$ \& P a base point on $E_q(a,b)$\\ \hline
$kP$ & Point multiplication in $E_q(a,b)$, where $k \in Z_q^{*}$ \\ \hline
$k_s$ \&  $Q_s$ & Master private key \& Public key of S, respectively, where $Q_s = k_sP$\\ \hline
$h(\cdot)$ & A collusion-resistant one-way hash function\\ \hline
$t_x$ & Current timestamp generated by entity $X \in \{U, S\}$\\ \hline
$||$ and $\oplus$ & Concatenation and Bit-wise XOR operation, respectively\\ \hline  
\end{tabular}
\end{center}
\end{table}

\section{Review and analysis of Sureshkumar et al.'s Scheme}
In this section, we briefly review Sureshkumar et al.  \cite{sureshkumar2017robust} proposed SIP and then present a security analysis. Sureshkumar et al. scheme consist of three phases, namely, system initialization, registration and authentication phases. The briefly review the three phases of Sureshkumar et al. scheme below. Note that hereafter we use the notations listed in Table \ref{notations}.

\small
\begin{tabular}{l}
\multicolumn{1}{c}{\textbf{Three phases of Sureshkumar et al.'s scheme}} \\ \\ \hline
\multicolumn{1}{c}{\textbf{Initialization phase}}\\ \hline
Select elliptic curve $E_q(a, b)$ and base point $P$ \\
Selects secure one-way hash function $h(\cdot)$\\
Selects master private key $k_s$ and compute public key $Q_s = k_sP$\\
Declares publicly $\{E_q(a,b), q, P, Q_s, h(\cdot)\}$\\ \hline
\end{tabular}

\resizebox{\textwidth}{!}{
\begin{tabular}{l l}
\multicolumn{2}{c}{\textbf{User registration phase}}\\ \hline
User $U$   & Server $S$ \\ \hline
Choose $ID_u$,$pw_u$ & \\ 
Computes $HIP_u = h(ID_u || pw_u)$,  $HID_u = h(ID_u)$,& \\
 $RP_u = ID_u \oplus pw_u$ &\\
$\xrightarrow[\hspace{1em}Private\ channel\hspace{1em}]{\hspace{1em}~~ \ Reg = \{HID_u, HIP_u, RP_u\}~~\hspace{1em}}$ & \\ 
				
    & Computes $UPW_u = h(HID_u || k_s) \oplus HIP_u$  \\
    &  Stores $\{UPW_u, RP_u\}$ in the database against $HID_u$. \\ \hline 
\end{tabular}
}

\resizebox{\textwidth}{!}{
\begin{tabular}{l l}
\multicolumn{2}{c}{\textbf{Login and key establishment phase}}\\ \hline
User $U$   & Server $S$ \\ \hline
Chooses $r_u \in Z_q^{*}$ & \\
Computes $R_u = r_uP$, $K_u = r_uQ_s$, & \\
$HID_u = h(ID_u)$, $HIP_u = h(ID_u|| pw_u)$, &\\ 
$D_u = HID_u \oplus h(K_u)$, $Auth_u = h(HIP_u || K_u || t_u)$ & \\
$\xrightarrow[\hspace{1em}Public\ channel\hspace{1em}]{\hspace{1em}~~ m_1 =\{R_u, D_u, Auth_u, t_u\}~~\hspace{1em}}$ &\\ 

     & Checks validity of $t_u$. Accept/Reject. \\
     & Computes $K_s = k_sR_u$, $HID_u = D_u \oplus h(K_s)$\\  
     & $HIP_u'$ = $UPW_u$ $\oplus h(HID_u || k_s)$\\ 
     & Checks $Auth_u =^? h(HIP_u' || K_s || t_u)$. Accept/Reject.\\
     & Selects $r_s \in Z_q^{*}$\\
     & Computes $R_s = r_sP$, $DK_s = r_sR_u$ \\ 
     & $Auth_s = h(HIP_u' || R_s || DK_s || t_s)$ \\
     & $SK_s = h(K_s || DK_s || HIP_u')$ \\
     & $\xleftarrow[\hspace{1em}Public\ channel\hspace{1em}]{\hspace{2em}~~m_2 =\{R_s, Auth_s, t_s\}~~\hspace{1em}}$ \\

Checks validity of $t_s$. Accept/Reject. & \\
Computes $DK_u = r_uR_s$ &\\
Checks $Auth_s =^? h(HIP_u || R_s || DK_u || t_s)$. Accept/Reject. & \\
Computes $SK_u = h(K_u || DK_u || HIP_u)$, & \\
$Conf = h(Auth_u || Auth_s || SK_u)$ &\\
$\xrightarrow[\hspace{1em}Public\ channel\hspace{1em}]{\hspace{1em}~~ m_3 =\{ Conf\} ~~\hspace{1em}}$  &\\ 
    
    &  Checks $Conf =^? h(Auth_u || Auth_s || SK_s)$. Accept/Reject. \\ \hline 
\end{tabular}
}

\subsection*{Security Analysis} In the following, we describe the security drawbacks of Sureshkumar et al.'s proposed scheme. Assume that an adversary captures all the transmitted messages between user U and server S via a public channel. The list of transcripts of each session are $\{m_1, m_2, m_3\}$, where $m_1 =\{R_u, D_u, Auth_u, t_u\}$, $m_2 =\{R_s, Auth_s, t_s\}$, and  $m_3 =\{ Conf\}$. Now we assume that the adversary launch session exposure attacks and get the session random secret $r_u \in Z_q^{*}$ \cite{canetti2001analysis}. The adversary $\mathcal{A}$ computes the user credentials $ID_u$ and $pw_u$ as follows using the revealed session ephemeral $r_u$ of user U:
\begin{itemize}
\item Computes $R_u^{*} = r_uP$. Checks whether $R_u^{*}$ matches with the parameter $R_u$ from $m_1$.
\item  If it matches, it guess the identity $ID_u$ as follows. Otherwise, repeat search for matching $R_u$.
\begin{itemize}
\item Compute $K_u = r_uQ_s$ and $HID_u = D_u \oplus h(K_u) = h(ID_u)$.
\item Guess an identity $ID_u^{*}$ and checks the validity of $HID_u = h(ID_u^{*})$. If valid guessed identity $ID_u^{*}$ is the original identity $ID_u$. 
\item Otherwise, repeat guessing until match. Since identity is chosen by user, off-line guessing of identity is not hard \cite{sureshkumar2017robust}. 
\end{itemize}
\end{itemize}

\par
\noindent Next adversary $\mathcal{A}$ launch the off-line password guessing attack as follows:
\begin{itemize}
\item Guess a password $pw_u^{*}$.
\item Check the validity of $Auth_u = h(h(ID_u || pw_u^{*}) || K_u || t_u)$. If it is valid, the guessed password is valid, that is, $pw_u^{*}$ is $pw_u$.
\item Otherwise, repeat the guessing until the match.
\end{itemize}

Therefore, from the above analysis, it is clear that Sureshkumar et al.'s proposed SIP fails to provide the user credentials privacy when the session-ephemeral secrets unexpectedly revealed to the adversary.

\section{Review and analysis of Zhang et al.'s Scheme}
In this section, we review and present the security analysis on Zhang et al.'s \cite{zhang2015secure}, and shows that their scheme also fail to protect the user credentials under CK-adversary assumption. Zhang et al.'s scheme consist three phases such as initialization, registration, and authentication phases. The initialization phase of Zhang et al.'s scheme is same as in Sureshkumar et al.'s scheme \cite{sureshkumar2017robust}. The other two phases registration and authentication of  Zhang et al.'s scheme are as follows. Note that user's realm is used to prompt the user identity and password.

\small
\begin{tabular}{l l}
\multicolumn{2}{c}{\textbf{User registration phase}}\\ \hline
User $U$   & Server $S$ \\ \hline
Chooses $ID_u, pw_u$   & \\
$\xrightarrow[\hspace{1em}Private\ channel\hspace{1em}]{\hspace{1em}~~ \{ID_u, pw_u\}~~\hspace{1em}}$ & \\ 

      & $VPW_u$=$h(ID_u || k_s)$$\oplus h(ID_u || pw_u)$ \\
      & Stores $\{ID_u, VPW_u\}$ in the database \\ \hline 
\end{tabular}

\resizebox{\textwidth}{!}{
\begin{tabular}{l l}
\multicolumn{2}{c}{\textbf{Login and key establishment phase}}\\ \hline
User $U$   & Server $S$ \\ \hline
Chooses $r_u \in Z_q^{*}$, & \\
Computes $R_u = r_uP$, $K_u = r_uQ_s$, & \\
$HID_u = ID_u \oplus h(R_u || K_u)$, & \\
$\xrightarrow[\hspace{1em}Public\ channel\hspace{1em}]{\hspace{1em}~~m_1 = \{HID_u, R_u\}~~\hspace{1em}}$ &\\ 

    & Chooses $r_s \in Z_q^{*}$, \\
    & Computes $R_s = r_sP$, $K_s = k_sR_u$, $DK_s=r_sR_u$, \\
    & $Auth_s = h(K_s || DK_s|| R_s || R_u)$\\ 
    & $\xleftarrow[\hspace{1em}Public\ channel\hspace{1em}]{\hspace{2em}~~m_2 = \{realm, R_s, Auth_s\}~~\hspace{1em}}$ \\

Computes $SK_u=abP$ & \\
Check $Auth_s =^? h(K_u || DK_u || R_s || R_u)$. Accept/Reject. & \\
Computes $SK = h(ID_u || DK_u || K_u || R_u || R_s)$, & \\
$Auth_u = h(realm || K_u || DK_u|| R_s || R_u || h(ID_u || pw_u))$ & \\
$\xrightarrow[\hspace{1em}Public\ channel\hspace{1em}]{\hspace{1em}~~m_3 = \{realm, Auth_u\}~~\hspace{1em}}$  &\\ 

     & Computes $ID_u' = HID_u \oplus h(R_u || K_s)$  \\
     &  $h(ID_u || pw_u) = VPW_u \oplus h(ID_u' || k_s)$\\
     &  Checks $Auth_u =^? h(realm || K_s || DK_s || R_s || R_u$ \\
     & $|| h(ID_u || pw_u))$. Accept/Reject.\\
     & Computes $SK=h(ID_u || DK_s || K_s || R_u || R_s)$\\ \hline 
\end{tabular}
}

\subsection*{Security Analysis}
We assume that an adversary captures all the transmitted messages between user $U$ and server $S$ via a public channel. The captured messages are $\{m_1, m_2, m_3\}$, where $m_1 = \{HID_u, R_u\}$ , $M_2 = \{realm, R_s, Auth_s\}$, and $m_3 = \{realm, Auth_u\}$. As defined, suppose the session ephemeral secret $r_u \in Z_q^{*}$ unexpectedly revealed to the adversary by the session exposure attacks \cite{canetti2001analysis}. Then the adversary $\mathcal{A}$ can compute the user credentials as follows:

\begin{itemize}
\item Computes $R_u^{*} = r_uP$ and $K_u^{*} = r_uQ_s$.
\item Checks whether $R_u^{*}$ matches with the parameter $R_u$ presented in $m_1$. 
\item If match found, $R_u = R_u^{*}$ and $K_u = K_u^{*}$ of user $U$ with identity $ID_u$. 
\item Computes user identity $ID_u = HID_u \oplus h(R_u || K_u)$. Then adversary launches off-line password guessing attack as follows.
\begin{itemize}
\item Computes $DK_u = r_uR_s$.
\item Guess password $pw_u^{*}$.
\item Checks the validity of $Auth_u = h(realm || K_u || DK_u || R_s || R_u || h(ID_u || pw_u^{*}))$. 
\item If it is valid, the guessed password $pw_u^{*}$ is original password $pw_u$. Otherwise, repeat the guessing until find the match.
\end{itemize}
\end{itemize}

From the above analysis, it is clear that Zhang et al.'s scheme fails to protect the user secret credentials $(ID_u, pw_u)$ when the session ephemeral secrets revealed to the adversary.

\section{Proposed Enhancement}
In this section, we propose an improvement over the Sureshkumar et al.'s scheme \cite{sureshkumar2017robust}. Our small modification in the storing parameters in the server database and small variation in the login request message, make the enhanced protocol is secure against the defined adversary. Our improved also have three phases, namely initialization, registration and authentication phases. The registration phase is same as in Sureshkumar et al.'s scheme \cite{sureshkumar2017robust}, and the other phases are presented below.

\resizebox{\textwidth}{!}{
\begin{tabular}{l l}
\multicolumn{2}{c}{\textbf{User registration phase}}\\ \hline
User $U$   & Server $S$ \\ \hline
Choose $ID_u$,$pw_u$ & \\ 
Computes $HIP_u = h(ID_u || pw_u)$,  $HID_u = h(ID_u)$,& \\
$\xrightarrow[\hspace{1em}Private\ channel\hspace{1em}]{\hspace{1em}~~ \ Reg = \{HID_u, HIP_u\}~~\hspace{1em}}$ & \\ 

     & Choose a random number $a_u \in Z_q^{*}$ \\
    & Computes $UPW_u = HID_u \oplus h(k_s || a_u) \oplus HIP_u$  \\
    &  Stores $\{UPW_u, a_u\}$ in the database against \textbf{$h(HIP_u || 1)$}. \\ \hline 
\end{tabular}
}

\resizebox{\textwidth}{!}{
\begin{tabular}{l l}
\multicolumn{2}{c}{\textbf{Login and key establishment phase}}\\ \hline
User $U$   & Server $S$ \\ \hline
Chooses $r_u \in Z_q^{*}$ & \\
Computes $R_u = r_uP$, $K_u = r_uQ_s$, & \\
$HIP_u = h(ID_u|| pw_u)$, &\\ 
$DP_u = HIP_u \oplus h(K_u)$, $Auth_u = h(HIP_u || K_u || t_u)$ & \\
$\xrightarrow[\hspace{1em}Public\ channel\hspace{1em}]{\hspace{1em}~~ m_1 =\{R_u, DP_u, Auth_u, t_u\}~~\hspace{1em}}$ &\\ 

     & Checks validity of $t_u$. Accept/Reject. \\
     & Computes $K_s = k_sR_u$, $HIP_u' = DP_u \oplus h(K_s)$\\  
     & Checks $Auth_u =^? h(HIP_u' || K_s || t_u)$. Accept/Reject.\\
     & Retrieve $UPW_u$ which is stored against $h(HIP_u' || 1)$\\
     & Chooses $r_s \in Z_q^{*}$\\
     & Computes $R_s = r_sP$, $DK_s = r_sR_u$ \\
     & $HID_u'$ = $UPW_u$ $\oplus h(k_s || a_u) \oplus HIP_u'$\\ 
     & $Auth_s = h(HID_u' || HIP_u' || R_s || DK_s || t_s)$ \\
     & $SK_s = h(K_s || DK_s || HIP_u' || HID_u')$ \\
     & $\xleftarrow[\hspace{1em}Public\ channel\hspace{1em}]{\hspace{2em}~~m_2 =\{R_s, Auth_s, t_s\}~~\hspace{1em}}$ \\

Checks validity of $t_s$. Accept/Reject. & \\
Computes $DK_u = r_uR_s$ &\\
Checks $Auth_s =^? h(HID_u || HIP_u || R_s || DK_u || t_s)$. Accept/Reject. & \\
Computes $SK_u = h(K_u || DK_u || HIP_u || HID_u)$, & \\
$Conf = h(Auth_u || Auth_s || SK_u || t_u || t_s)$ &\\
$\xrightarrow[\hspace{1em}Public\ channel\hspace{1em}]{\hspace{1em}~~ m_3 =\{ Conf\} ~~\hspace{1em}}$  &\\ 
    
    &  Checks $Conf =^? h(Auth_u || Auth_s || SK_s || t_u || t_s)$. Accept/Reject. \\ \hline 
\end{tabular}
}

\subsection*{Security Analysis}
In our improved version of the protocol, we updated the server database table as $h(HIP_u || 1)$ : $\{UPW_u, a_u\}$ for user $U$. We then modified the request message as $m_1 =\{R_u, DP_u, Auth_u, t_u\}$. In our scheme, we are sending $DP_u = HIP_u \oplus h(K_u)$ instead of sending $D_u = HID_u \oplus h(K_u)$ in the Sureshkumar et al.'s scheme \cite{sureshkumar2017robust}. In our case, even if the session ephemeral secret revealed to an adversary $\mathcal{A}$, he/she can only retrieve $HIP_u = h(ID_u || pw_u)$. Therefore, guessing both identity $ID_u$ and password $pw_u$ simultaneously makes hard to the adversary than guessing one-by-one individually the identity $ID_u$ and password $pw_u$. Whereas, other protocols leave the option of guessing individually the identity and the password. Therefore, our improved version provides strong credential privacy even in the case of ephemeral leakage without increasing the computational overheads over the Sureshkumar et al.'s protocol \cite{sureshkumar2017robust} and Zhang et al.'s protocol \cite{zhang2015secure}.

\section{Performance analysis}
In this section, we discuss the performance comparison as well as the security features satisfied by various related schemes in the literature.

\begin{table}[!htp]
\small
\begin{center}
\caption{Computation cost comparison} \label{compare-cost}
\begin{tabular}{|l|p{2cm} p{2cm}|p{2cm}|p{2cm}|}
\hline
\multirow{2}{*}{Scheme}  & \multirow{1}{*}{Participant}  &   &  \multirow{2}{*}{Total overhead}  & \multirow{2}{*}{Running Time} \\ \cline{2-3} 
       
       & User U & Server S   &  &   \\ \hline

Lu et al. \cite{lu2017anonymous}    & 3$T_{pm}$+8$T_h$   &  3$T_{pm}$+7$T_h$     &  6$T_{pm}$+15$T_h$  &  0.1074 sec   \\  \hline

Chaudhry et al. \cite{chaudhry2017improved}    & 3$T_{pm}$+5$T_h$   &  3$T_{pm}$+5$T_h$     &  6$T_{pm}$+10$T_h$  &  0.1058 sec   \\  \hline

Tu et al. \cite{tu2015improved}  &   $3T_{pm}$+5$T_h$  &  $3T_{pm}$+5$T_h$     &   $6T_{pm}$+10$T_h$    & 0.1058 sec   \\  \hline


Farash \cite{farash2016security}   & 4$T_{pm}$+5$T_h$   &  3$T_{pm}$+5$T_h$     &  7$T_{pm}$+10$T_h$  &  0.1229 sec   \\  \hline

Lu et al. \cite{lu2016secure}  &  2$T_{pm}$+4$T_h$   &  2$T_{pm}$+5$T_h$     &   4$T_{pm}$+9$T_h$      &  0.07128 sec \\  \hline

Arshad et al. \cite{arshad2016efficient} &   2$T_{pm}$+4$T_h$  &  2$T_{pm}$+4$T_h$     &   $4T_{pm}$+8$T_h$  &  0.07096 sec  \\  \hline 


Zhang et al. \cite{zhang2015secure}   &  3$T_{pm}$+4$T_h$  &  3$T_{pm}$+5$T_h$     &   6$T_{pm}$+9$T_h$  & 0.10548 sec \\  \hline

Sureshkumar et al. \cite{sureshkumar2017robust}   & 3$T_{pm}$+7$T_h$   &  3$T_{pm}$+5$T_h$     &  6$T_{pm}$+12$T_h$  &  0.10644 sec   \\  \hline

Ours   & 3$T_{pm}$+6$T_h$   &  3$T_{pm}$+5$T_h$   &  6$T_{pm}$+11$T_h$  &  0.10612 sec   \\  \hline
\end{tabular}
\end{center}
\end{table}

\par
In Table \ref{compare-cost}, we compare the computational overheads required in various existing protocols as well as our improved version of the protocol. To analyze the computational cost, we use the notations for different cryptographic operations as follows: $T_{pm}$ : ECC point multiplication and $T_h$: cryptographic hash function, and we omit the other lightweight operation such as symmetric-key encryption/decryption and bitwise exclusive-OR operations in our comparison. In order to estimate the approximate execution timings, we use the experimental results presented in He et al.'s work \cite{he2014enhanced}. The approximate execution timings are $T_{pm}$ $\approx$ 0.0171sec and $T_h$ $\approx$ 0.00032sec. From the Table \ref{compare-cost}, it is clear that our proposed improvement requires little lesser computational cost compared to the original  Sureshkumar et al.'s protocol \cite{sureshkumar2017robust}, and our improvement is also comparable with the other existing protocols.

\begin{table}[!htp]
\begin{center}
\caption{Security requirement comparison} \label{compare-security}
\resizebox{\textwidth}{!}{
\begin{tabular}{|l|c|c|c|c|c|c|c|c|c|c|c|}
\hline
Scheme   & $~~~Z_1~~~$  & $~~~Z_2~~~$    & $~~~Z_3~~~$     & $~~~Z_4~~~$     & $~~~Z_5~~~$      & $~~~Z_6~~~$       & $~~~Z_7~~~$      & $~~~Z_8~~~$        & $~~~Z_9~~~$      & $~~~Z_{10}~~~$        & $~~~Z_{11}~~~$      \\ \hline

Lu et al. \cite{lu2017anonymous}     &  N  &  N  &  Y    &  Y   & N   &  Y   &  Y    &  Y   &  Y  &  Y  &  N  \\  \hline
			
Chaudhry et al. \cite{chaudhry2017improved}              &    N  &  Y   &  Y   &  N   &  Y   & Y   &  Y    &  Y   & Y   &  Y   & N  \\  \hline
			
Tu et al. \cite{tu2015improved}       &  N &  Y  & N  & Y  &  N  &  N  &  Y   &  Y   & Y   & Y   & N \\  \hline


Farash \cite{farash2016security} &  N  &  N  & N    & Y   & N  & N  & Y   &  Y  &  Y  &  N  & N \\  \hline

Lu et al. \cite{lu2016secure}         &  N  & Y  & N & Y  & N & N  &  Y   &  Y  &  Y  & Y   & N  \\  \hline

Arshad et al. \cite{arshad2016efficient}   &  N & Y   & Y  & Y  & N  & Y & Y & Y  & Y  & Y  & N \\   \hline


Zhang et al. \cite{zhang2015secure}    &  Y  &  Y  &  N   &  N   &  N   &  N   &  Y   &  Y  &  Y   & Y  & N  \\  \hline

Sureshkumar et al. \cite{sureshkumar2017robust}   & Y  &  N   & Y   & Y  & Y  & Y   & Y   & Y   & Y   &  Y  & N  \\  \hline

Ours  & Y & Y & Y  & Y & Y & Y  & Y & Y & Y & Y & Y \\ \hline
\end{tabular}
}
\end{center}
\textit{Note:}	 
$Z_1$: Achieves user anonymity; 
$Z_2$: Withstand off-line password guessing attack;
$Z_3$: Withstand impersonation attack; 
$Z_4$: Withstand insider attack; 
$Z_5$: Withstand replay attack;
$Z_6$: Achieves strong mutual authentication; 
$Z_7$: Withstand stolen verifier attack; 
$Z_8$: Provides session key security;
$Z_9$: Achieves perfect forward secrecy; 
$Z_{10}$: Withstand man-in-the-middle attack;
$Z_{11}$: Whether provide credentials privacy when session ephemeral revealed to an adversary; \\
Y : Provides the security feature; N : Does not provide the security feature.	
\end{table}

In the Table \ref{compare-security}, we compare security features satisfied by the various related existing protocols  \cite{lu2017anonymous,chaudhry2017improved,tu2015improved,farash2016security,lu2016secure,arshad2016efficient,zhang2015secure,sureshkumar2017robust} with our improved version of the protocol. We can observe that Lu et al. \cite{lu2017anonymous}, Chaudhry et al. \cite{chaudhry2017improved}, Tu et al. \cite{tu2015improved}, Farash \cite{farash2016security}, Lu et al. \cite{lu2016secure} and Arshad et al. \cite{arshad2016efficient} failed to provide user anonymity because most of them send his/her username/identity in plaintext to the server through a public channel or we can compute the identity easily from the transmitted messages. However, according to our observation, from the Table \ref{compare-security},  Lu et al. \cite{lu2017anonymous}, Tu et al. \cite{tu2015improved}, Farash \cite{farash2016security}, Lu et al. \cite{lu2016secure}, Arshad et al. \cite{arshad2016efficient} and Zhang et al. \cite{zhang2015secure} schemes fail to resist from replay attack. Off course we can prevent this attack by properly merging the timestamp in the transmitted messages. In addition, the impersonation attacks is also a serious concern where Tu et al. \cite{tu2015improved}, Farash \cite{farash2016security}, Lu et al. \cite{lu2016secure},and Zhang et al. \cite{zhang2015secure} fails to provide it. Our improved version of the protocol provides more security features along with the comparable computational overheads compared to the other existing schemes in the literature. As a result, our proposed improvement outperform in terms of computational efficiency along with offers increased security features.

\section{Conclusion and Future Work}
We have first analyzed the security limitations of the recently proposed Sureshkumar et al. and Zhang et al.'s session initiation protocols. We have shown that both the schemes fail to protect the user secret credentials (identity and password) when the session ephemeral secrets are unexpectedly revealed to an adversary by the session exposure attacks. The presented security analysis in this paper is also applicable to most of the existing schemes in the literature. We then discuss the possible improvement to overcome the pitfalls find the existing schemes. In addition, we present the security and performance comparisons of the related existing schemes in the literature and compare with the proposed enhanced scheme. In our observation, the further study is required in this area of research to design secure and efficient session initiation protocols for quality healthcare services. In the future work, we aim to explore novel privacy preserving approaches for session initiation protocol, particular to the emergency healthcare application.

%
%

\bibliographystyle{unsrt}
\bibliography{SSCC2018}

%
%
%
%
%

\end{document}